\documentclass[aps,12pt,superscriptaddress,twocolumn,nofootinbib,tightenlines]{revtex4}
\usepackage{natbib}
\usepackage{graphicx}
\usepackage{dcolumn}
\usepackage{amsmath}
\usepackage{amssymb}

\renewcommand{\le}{\leqslant}
\renewcommand{\leq}{\leqslant}

\newcommand{\E}{\mbox{\textsf{E}}}        
\newcommand{\w}[2]{\ensuremath{w_{#1}^{(#2)}}}
\newcommand{\D}{\displaystyle}
\renewcommand{\r}[2]{\ensuremath{r_{#1}^{(#2)}}}

\newcommand{\remark}[1]{}
\renewcommand{\remark}[1]{#1}

\begin{document}

\title{A simple evolutionary game with feedback between perception and reality}
\author{Dmitriy Cherkashin}
\affiliation{Department of Mathematics, University of Chicago}
\author{J.~Doyne Farmer}
\affiliation{Santa Fe Institute}
\author{Seth Lloyd}
\affiliation{Department of Mechanical Engineering, MIT}

\begin{abstract}
We study an evolutionary game of chance in which the probabilities
for different outcomes (e.g., heads or tails) depend on the amount
wagered on those outcomes.  The game is perhaps the simplest
possible probabilistic game in which perception affects reality. By
varying the `reality map', which relates the amount wagered to the
probability of the outcome, it is possible to move continuously from
a purely objective game in which probabilities have no dependence on
wagers, to a purely subjective game in which probabilities equal the
amount wagered.  The reality map can reflect self-reinforcing
strategies or self-defeating strategies.  In self-reinforcing games,
rational players can achieve increasing returns and manipulate the
outcome probabilities to their advantage; consequently, an early
lead in the game, whether acquired by chance or by strategy,
typically gives a persistent advantage.  We investigate the game
both in and out of equilibrium and with and without rational
players.  We introduce a method of measuring the inefficiency of the
game, and show that the inefficiency decreases slowly in the
approach to equilibrium (for large $t$ it is a power law $t^{-\gamma}$ with $0 \le
\gamma \le 1$, depending on the subjectivity of the game).
\end{abstract}

\maketitle


\section{Introduction}

\subsection{Motivation}

To capture the idea that objective outcomes depend on subjective
perception Keynes used the metaphor of a beauty contest in
which the goal of the judges is not to decide who is most beautiful,
but rather to guess which contestant will receive the most votes from the other
judges \cite{Keynes36}.  Economic problems
typically have both purely objective components, e.g., how much
revenue a company creates, as well as subjective components, e.g.,
how much revenue investors collectively {\it think} it will create.  The
two are inextricably linked: subjective perceptions alter investment
patterns, which affect objective outcomes, which in turn affect
subjective perceptions.  

To study this problem this paper
introduces a simple probabilistic game in which the probability of
outcomes depends on the amount bet on those outcomes.  
The game introduces the idea of a `reality map' that mediates
between subjective perception and objective outcome.  This goes beyond Keynes,
in that subjective perception actually alters the reality (in his example the faces of
the contestants).
The form of
the reality map can be tuned to move continuously from purely
objective to purely subjective games, and different levels of
feedback from perception to reality are easily
classified and studied.\footnote{The results in this paper are
presented in more detail in reference \cite{Cherkashin04}.}

Consider a probabilistic event, such as a coin toss or the outcome
of a horse race.  Now suppose that the odds of the outcomes of the
event depend on the amount wagered on them.  In the case of a coin
toss, this means that the probability of heads is a function (the
`reality map') of the amount bet on heads.  For a purely objective
event, such as the toss of a fair coin, the reality map is simple:
the probability of heads is 1/2, independent of the amount bet on
 it.  But most events are not fully objective.  In the case of a horse
race, for instance, a jockey riding a strongly favored horse may
make more money if he secretly bets on the second most favored horse
and then intentionally loses the race.  This is an example of a
 self-defeating map from perception to reality: if jockeys misbehave, then as the horse becomes
more popular, the objective probability that it will win decreases.
Alternatively, in an economic setting, if people like growth
strategies they will invest in companies whose prices are going up,
which in turn drives prices further up.  This is an example of a
self-reinforcing reality map.

\subsection{Review of related work}

There has been considerable past work on situations where subjective
factors influence objective outcomes.  Some examples include
Hommes's studies of cobweb models \cite{Hommes91,Hommes94}, studies of increasing
returns \cite{Arthur94:book}, Arthur's
El Farol model and its close relative the
minority game \cite{Arthur94:paper,Challet05}, Blume and Easley's model of the
influence of capital markets on natural selection in an economy
\cite{Blume-Easley92,Blume-Easley02}, and Akiyama and Kaneko's
example of a game that changes due to the players' behaviors and
states \cite{akiyama-kaneko00}.  The model we introduce here has the
advantage of being very general yet very simple, providing a tunable way to study this phenomenon under varying levels of feedback.

\section{Game definition}

\subsection{Wealth dynamics}

Let $N$ agents place wagers on $L$ possible outcomes.  In the case
of betting on a coin, for example, there are two outcomes, heads and
tails. Let $s_{il}$ be the fraction of the $i$-th player's wealth
$w_i$ that is wagered on the $l$-th outcome.  The vector
 $(s_{i1},\dots,s_{iL})$ is the $i$-th player's \emph{strategy},
and $p_{il} = s_{il} w_i$ is the amount of money bet on the $l$-th
outcome by player $i$.  Let $p_l=\sum_i p_{il}$ be the total wager
 on the $l$-th outcome.  If the winning outcome is $l =  \lambda$,
the payoff $\pi_{i\lambda}$ to player $i$ is proportional to the
amount that player bets and inversely proportional to the total
amount everyone bets, i.e.
\[
\pi_{i\lambda} = \frac{p_{i\lambda}}{p_\lambda} =
\frac{s_{i\lambda}w_i}{p_\lambda}\,.
\]
This corresponds to what is commonly called pari-mutuel betting.  We
assume no ``house take", i.e.\ a fair game.  Assume \ $\sum_l s_{il} =
1$, i.e. that each player bets all her money at every iteration of the game, typically betting non-zero amounts on each possible outcome.
The total wealth is conserved and is normalized to sum to one,
\[
\sum_i w_i=\sum_{i,l} p_{il} = 1\,.
\]

We will call $q_l$ the probability of outcome $l$, where $\sum_l q_l
= 1$.  The expected payoff $\E \pi_i$ is
\[
\E \pi_i = \sum_l q_l \pi_{il}\,.
\]
 If the vector $q$ is fixed, after playing the game repeatedly for $t$ rounds
the wealth updating rule
\[
w_i^{(t+1)}=\frac{s_{i\lambda}w_{i}^{(t)}}{p_\lambda}
\]
is equivalent to Bayesian inference, where the initial wealth
 $w_i^{(t)}$ is interpreted as the prior probability that $q_\lambda =
s_{i\lambda}$ and the final wealth $w_i^{(t+1)}$ is its posterior
probability.  In Bayesian inference, models whose predictions match
the actual probabilities of outcomes accrue higher \textit{a
posteriori} probability as more and more events occur.  Here
players whose strategies more closely match actual outcome
probabilities accrue wealth on average at the expense of players
whose strategies are a worse match.

\subsection{Strategies}

We first study fixed strategies.  For convenience we restrict the
possible number of outcomes to $L = 2$, so that we can think of this
as a coin toss with possible outcomes heads and tails.  Because the
players are required to bet all their money on every round,  $s_{i1}
+ s_{i2} = 1$, we can simplify the notation and let $s_i = s_{i1}$
be the amount bet on heads --- the amount bet on tails is determined
 automatically.  Similarly $q = q_1$ and $p = p_1$. The space of possible strategies corresponds to the
unit interval $[0,1]$. We will typically simulate $N$ fixed
strategies, $s_i = i/(N - 1)$, where $i = 0, 1, \ldots, N-1$.
 Later
on we will also add rational players, who know  the strategies of
 all other players and dynamically adapt their own strategies to maximize
a utility function.

\subsection{Reality maps}

The game definition up to this point follows the game studied by
 Cover and Thomas \cite{CoverThomas:book}.  We generalize their game by
allowing for the possibility that the objective probability $q$ for
heads is not fixed, but rather depends on the net amount $p$ wagered
on heads. The \textit{reality map} $q(p)$, where $0 \leq q(p) \leq
1$, fully describes the relation between bets and outcomes.  We
restrict the problem slightly by requiring that $q(1/2) = 1/2$.  We
do this to give the system the chance for the objective outcome, as
manifested by the bias of the coin, to remain constant at $q = 1/2$.
We begin by studying the case where $q(p)$ is a monotonic function,
which is either nondecreasing or nonincreasing.  Letting $q'(p) =
dq/dp$, we distinguish the following possibilities:
\begin{itemize}
\item \textit{Objective}.  $q(p) = 1/2$,  i.e.\ it is a fair coin independent of the amount wagered.  (Other
values of $q = \text{constant}$ are qualitatively similar to $q=1/2$.)
\item \textit{Self-defeating.}  $q'(p) < 0$, e.g.\ $q(p) = 1 - p$. In this case the coin tends to oppose the collective perception, e.g.\ if people collectively bet on heads, the coin is biased toward tails.
\item \textit{Self-reinforcing.}  $q'(p) > 0$. The coin tends to reflect the collective perception, e.g.\ if people collectively bet on heads, the coin becomes more biased toward heads.  A special case of this is \textit{purely subjective}, i.e. $q(p) = p$, in which the bias simply reflects people's bets.
\end{itemize}

It is convenient to have a one parameter family of reality maps that
allows us to tune from objective to self--reinforcing. We choose the
family
\begin{equation}
\label{qalpha}
q_{\alpha}(p) = \frac{1}{2} + \frac{1}{\pi} \arctan \frac{\pi \alpha
(p-\frac{1}{2})}{1-(2p-1)^2}.
\end{equation}
The parameter $\alpha$ is the slope at $p = 1/2$.  When $\alpha =
0$, $q(p)$ is constant (purely objective), and when $\alpha > 0$,
 $q(p)$ is self-reinforcing.  The derivative $q'_\alpha(1/2)$ is an
 increasing function of $\alpha$; when $\alpha = 1$, $q'(1/2) = 1$, and
 $q(p)$ is close to the identity map.\footnote{An inconvenient aspect of
this family is that $q_{\alpha}(p)$ does not contain the function
$q(p)=p$.  However, $q_1(p)$ is very close to $q(p)=p$ (the
difference does not exceed $0.012$, with the average value less than
 half of this).  Still, to avoid any side effects, we study the purely subjective case using
$q(p) = p$.}  We study the self-defeating case separately using the
map $q(p) = 1 - p$.

\begin{figure}[htb]
\begin{center}
\includegraphics[scale = 0.5]{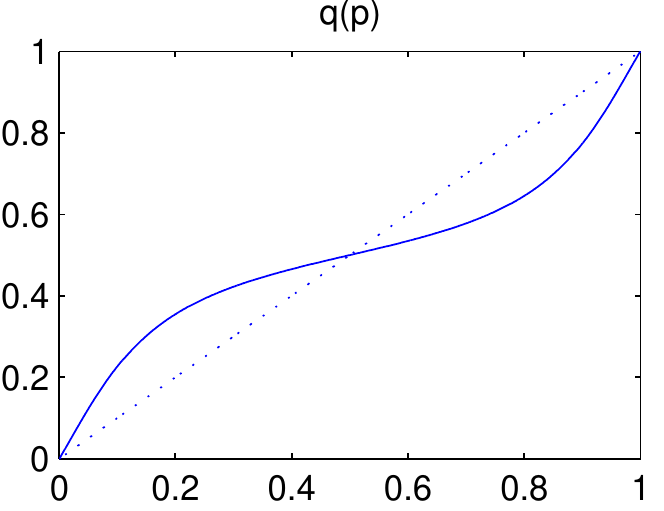}
\includegraphics[scale = 0.5]{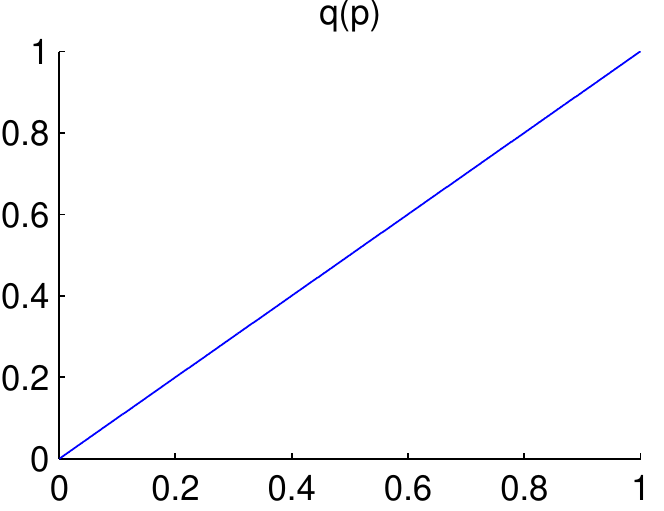}
\includegraphics[scale = 0.5]{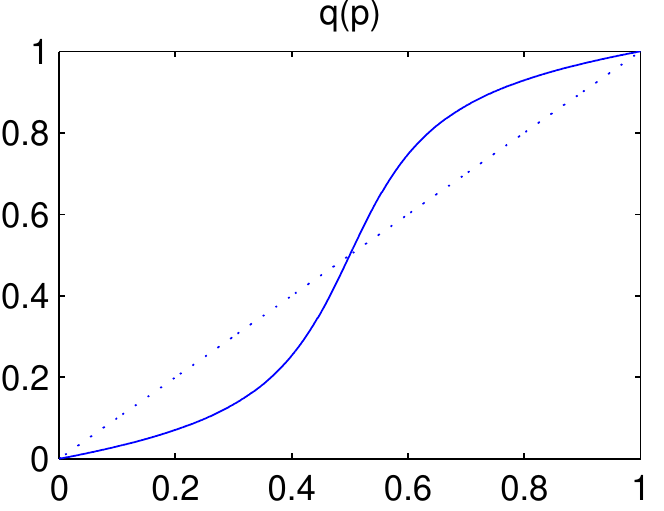}
\caption{$q_\alpha(p)$ with different values of parameter: (a)
$\alpha=1/3$; (b) $\alpha=1$; (c) $\alpha=3$.}
\end{center}
\end{figure}

\section{Dynamics of the objective bias}

\begin{figure}[htb]
\begin{center}
\includegraphics[scale = 0.2]{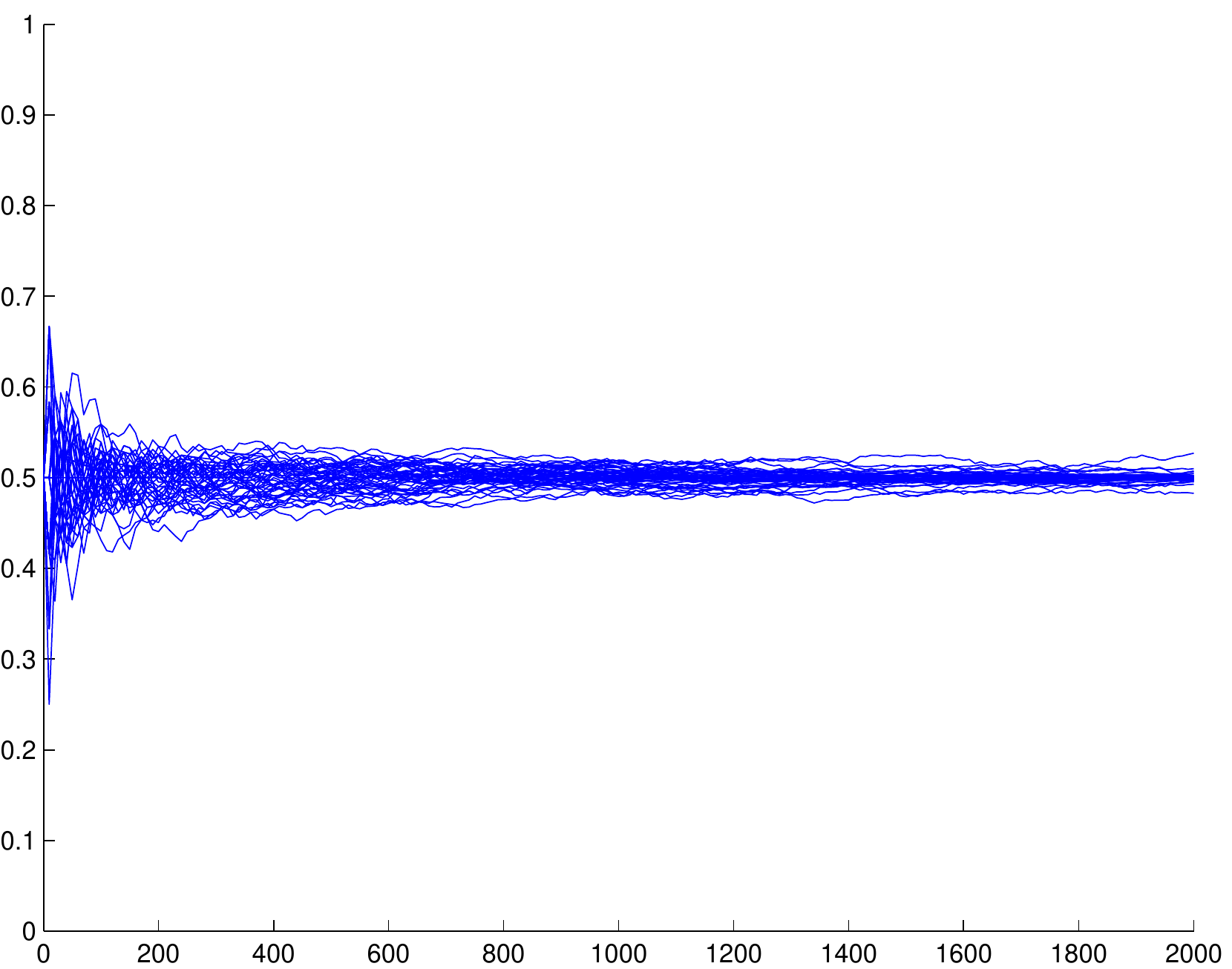}
\includegraphics[scale = 0.2]{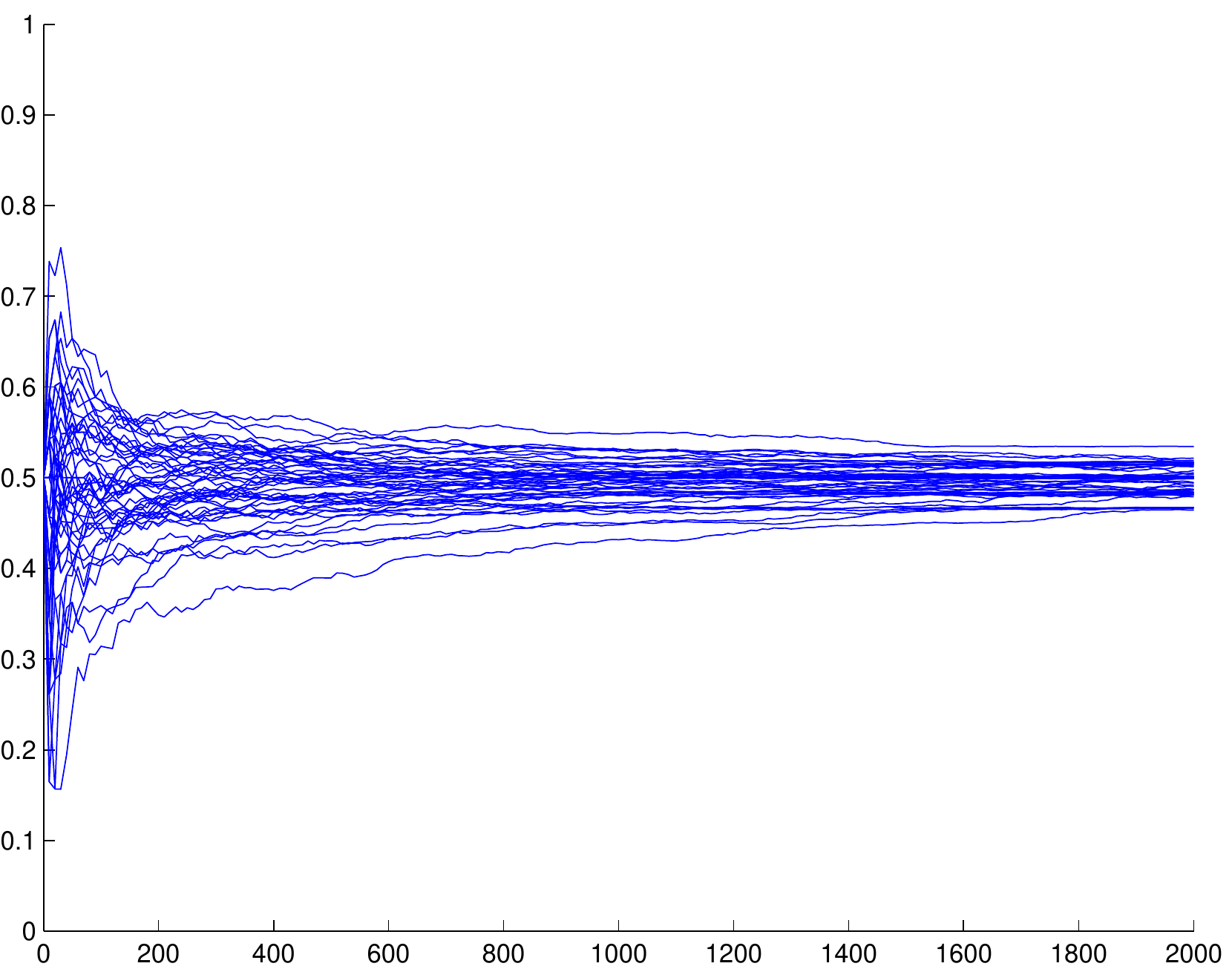}
\includegraphics[scale = 0.2]{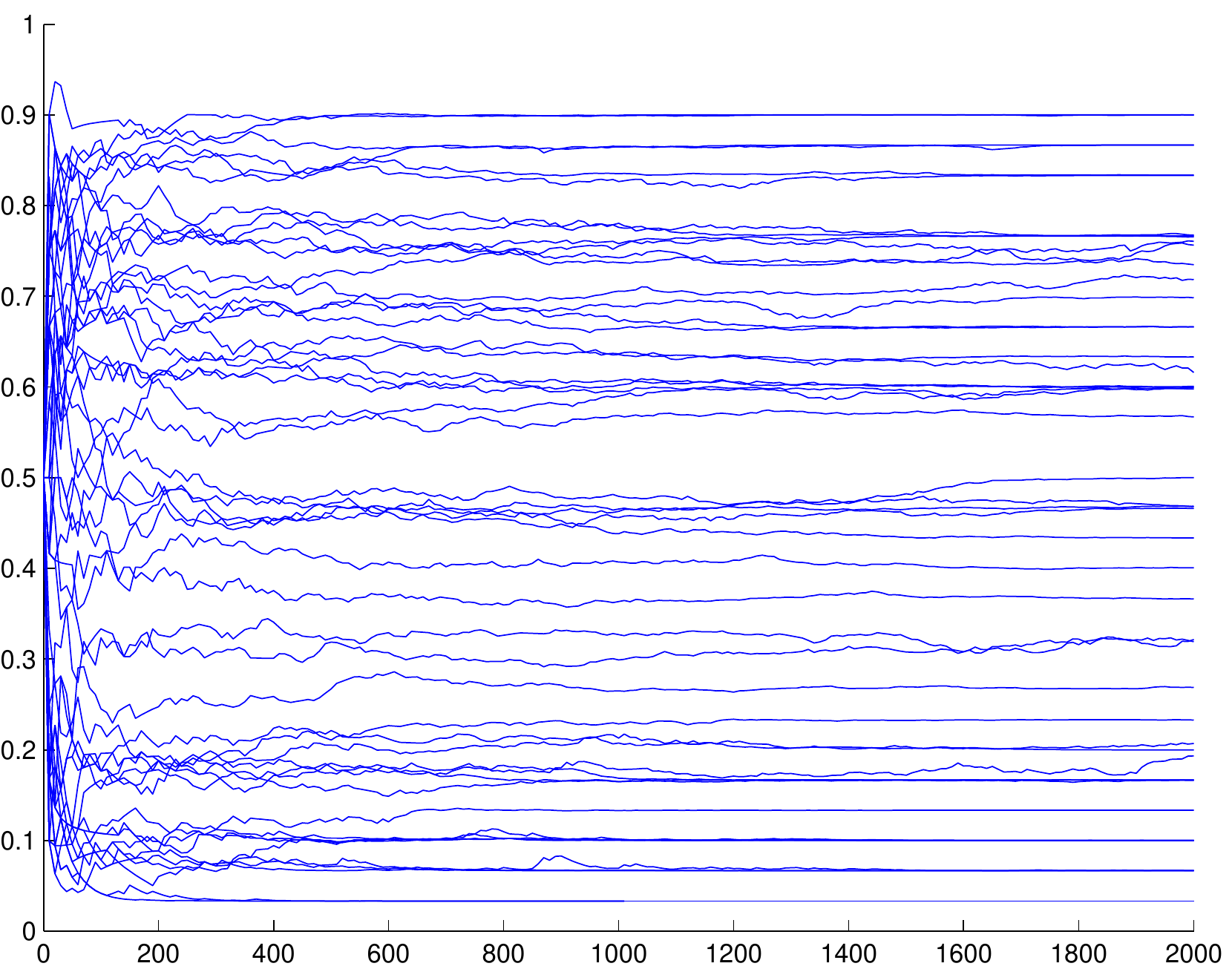}
\includegraphics[scale=0.2]{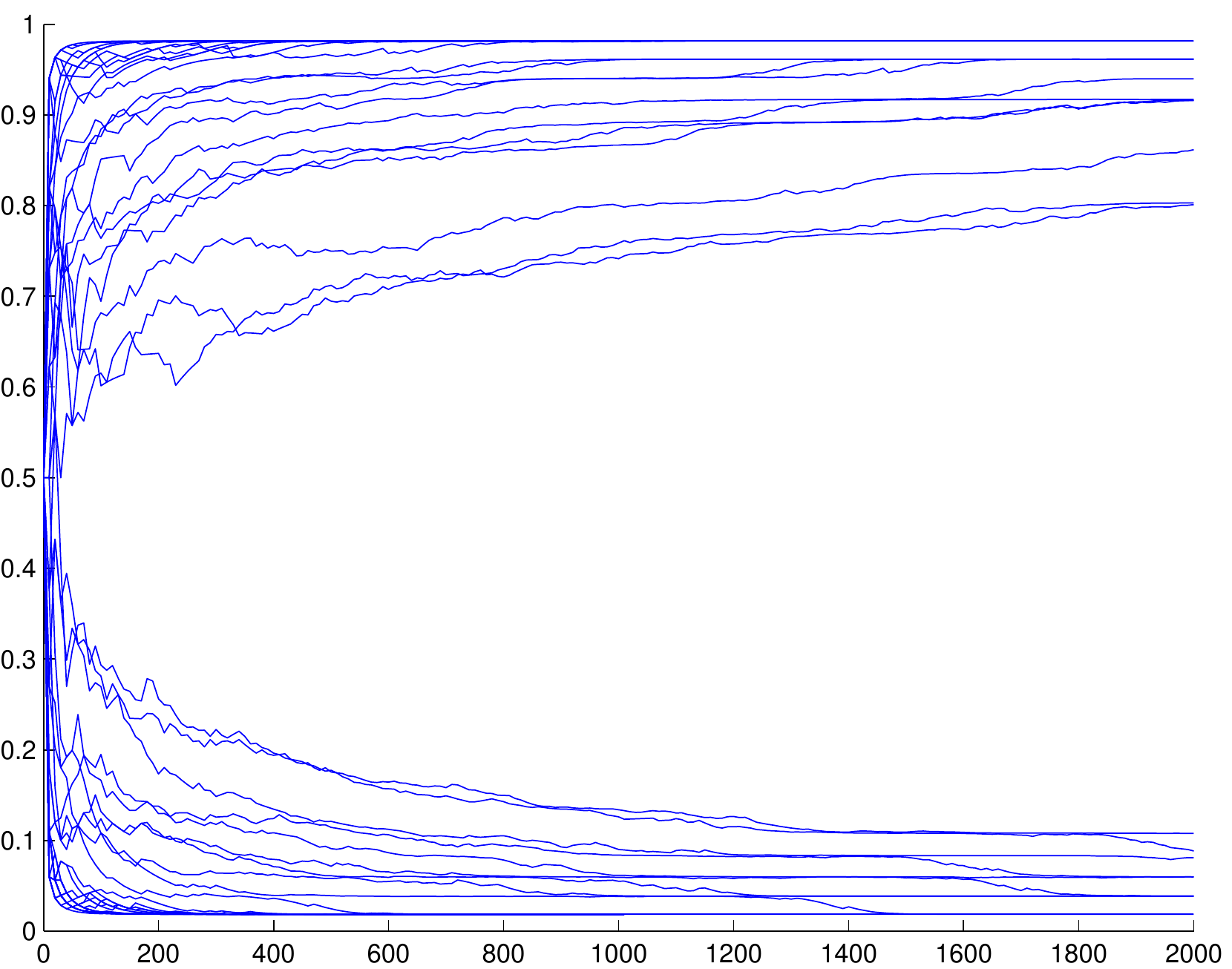}
\includegraphics[scale=0.2]{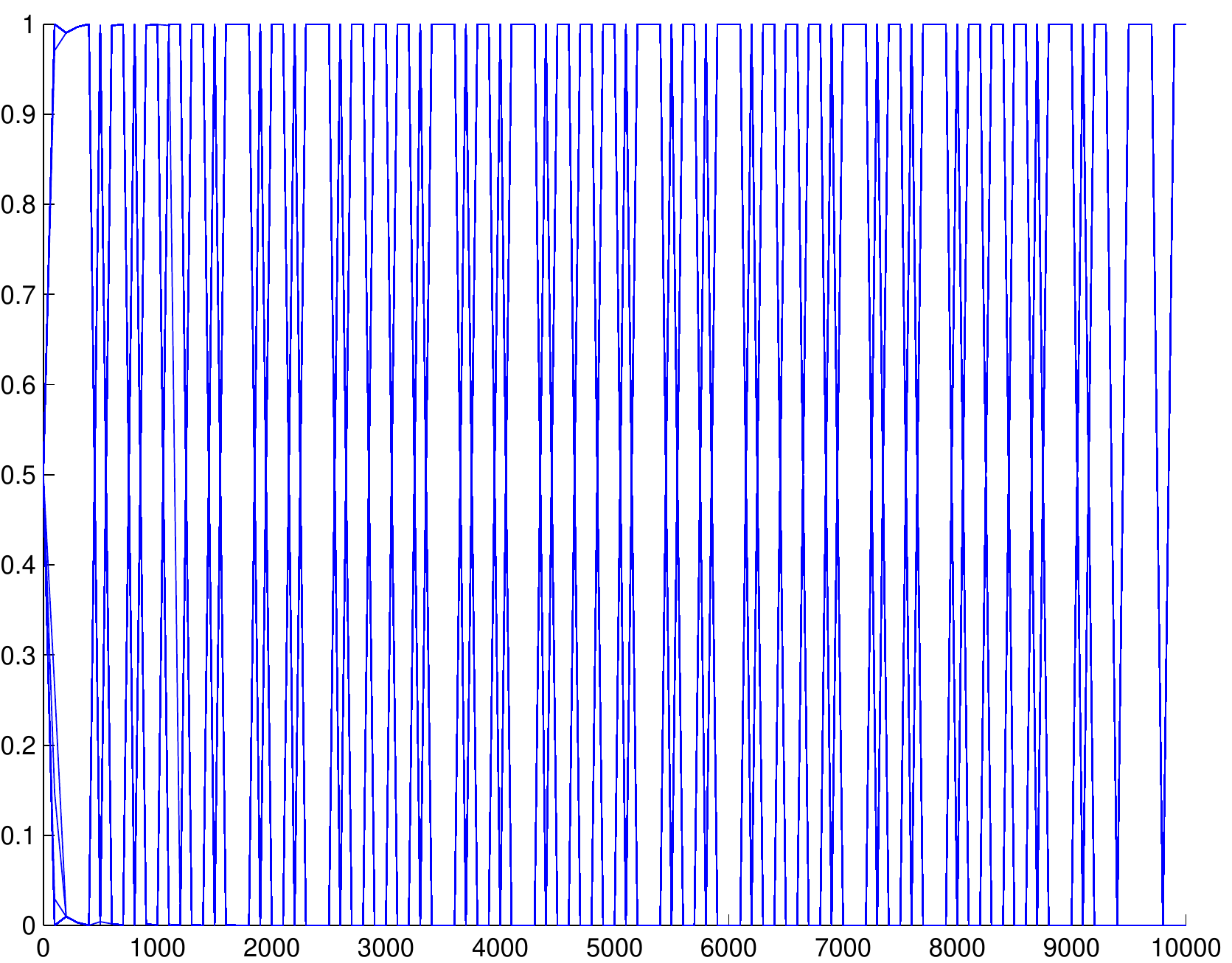}
\caption{The objective bias of the coin $q^{(t)}$ as a function of
time, for different random number seeds.  (a) q(p) = 1 - p; (b) in
equation~\eqref{qalpha}, $\alpha = 1/2$; (c) q(p) = p; (d) $\alpha =
1.5$;  (e) $q(p) = 3p \mod 1$. (a)--(d) 50 runs up to $t=2000$; (e)
100 runs up to $t=10000$.} \label{biasDynamics}
\end{center}
\end{figure}

In this section we study the dynamics of the objective bias of the
coin, which is the tangible reflection of ``reality'' in the game.
This also allows us to get an overview of the behavior for different
 reality maps $q(p)$.  We use $N= 29$ agents each playing one of the 29 equally spaced strategies on $(0,1)$: $1/30, 2/30, \dots, 29/30$, and begin by giving
them all equal wealth.  We then play the game repeatedly and plot
the bias of the coin $q^{(t)}$ as a function of time.  This is done
several times to get a feeling for the variability vs.\ consistency
of the behavior of different reality maps, as shown in
Figure~\ref{biasDynamics}.

For the purely objective case $q(p) =
1/2$, the result is trivial.  For the self-defeating case,  $q(p) =
1 - p$, the results become more interesting, as shown in
(a). Initially the bias of the coin varies considerably, with a range that is
 generally about $0.3$ -- $0.7$, but it eventually settles into a fixed
point at $q = 1/2$.  For this case the bias tends to
oscillate back and forth as it approaches its equilibrium value. Suppose, for example, that the
first coin toss yields heads; after this toss, players who bet more
on heads possess a majority of the wealth.  At the second toss,
because of the self-defeating nature of the map, the coin is biased
towards tails.  As a result, wealth tends to shift back and forth
between heads and tails players before finally accruing to players
who play the `sensible' unbiased strategy.

We then move to the weakly self-reinforcing case using
equation~\eqref{qalpha} with $\alpha = 1/2$, as shown in (b).  The
behavior is similar to the previous case, except that the
 fluctuations of $q^{(t)}$ are now larger.  At the end of $2000$ rounds
of the game, the bias is much less converged on $q = 1/2$.  The bias
 is also strongly autocorrelated in time --- if the bias is high at a given
time, it tends to remain high at subsequent times.  (This was
already true for the self-defeating case, but is more pronounced
here). Although this is not obvious from this figure, after a
sufficiently long period of time all trajectories eventually
converge to $q = 1/2$.

Next we study the purely subjective case, $q(p) = p$, as shown in
(c).  In this case the bias fluctuates wildly in the early rounds of
the game, but it eventually converges to one of the strategies
$s_i$, corresponding to the player who eventually ends up with all
the wealth.

As we increase $\alpha > 1$,  as shown in (d), the instability
becomes even more pronounced.  The bias initially fluctuates near $q
 = 1/2$, but it rapidly diverges to fixed points either at $q
= 0$ or $q = 1$.  Which of the two fixed points is chosen depends on
the random values that emerge in the first few flips of the coin;
initially the coin is roughly fair, but as soon as a bias begins to
develop, it is rapidly reinforced and it locks in.  The extreme case
occurs when $q(p)$ is a step function, $q(p) = 0$ for $0 \le p <
 1/2$, $q(1/2)=1/2$ and $q(p) = 1$ for $1/2 < p \le 1$.  In this case the first
coin flip determines the future dynamics entirely: If the first coin
flip is heads, then players who favor heads gain wealth relative to
those who favor tails, and the coin forever after yields heads,
until all the wealth is concentrated with the player that bets most
heavily on heads.  (And vice versa for tails).

Finally in (e) we show an example of the bias dynamics for the
multi-modal map $q(p) = 3p \mod 1$.  In this case the bias
oscillates between $q = 0$ or $q = 1$, with a variable period that
is the order of a few hundred iterations.  We explain this behavior
at the end of the next section.

\section{Wealth dynamics\label{wealthDynamics}}

How do the wealths of individual players evolve as a function of
time?  The purely objective case $q = \text{constant}$ with fixed
 strategies and using a bookmaker instead of pari-mutuel betting was studied by Kelly \cite{kelly:gambling} and summarized by Cover and Thomas \cite{CoverThomas:book}.  Assuming
all the strategies are distinct, they show that the agent with the
strategy closest to $q$ asymptotically accumulates all the wealth.
Here ``closeness" is defined in terms of the
Kullback-Leibler distance between the strategy $s_{il}$ and the true
 probability $q_l$.

 For all reality maps $q(p)$ that we have studied we find
that one player asymptotically accumulates nearly all the wealth.
As a particular player becomes more wealthy, it becomes less and
less likely that another player will ever overtake this player.
 This concentration of wealth in the
hands of a single player is the fundamental fact driving the
convergence of the objective bias dynamics to a fixed point, as
observed in the previous section.  The reason is simple:  once one
player has all the wealth, this player completely determines the
odds, and since her strategy is fixed, she always places the same
bets.

\begin{figure}[htbp]
\begin{center}
\includegraphics[width=0.23\textwidth]{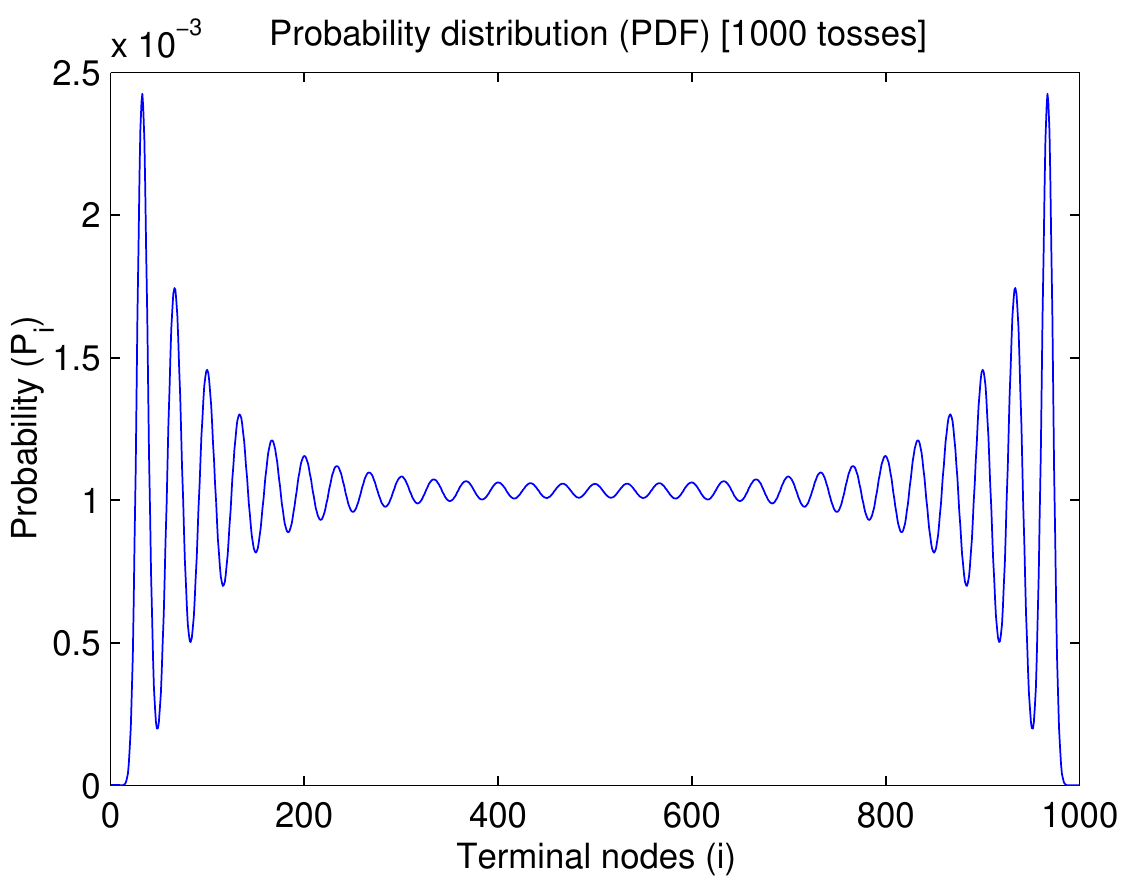}
\includegraphics[width=0.23\textwidth]{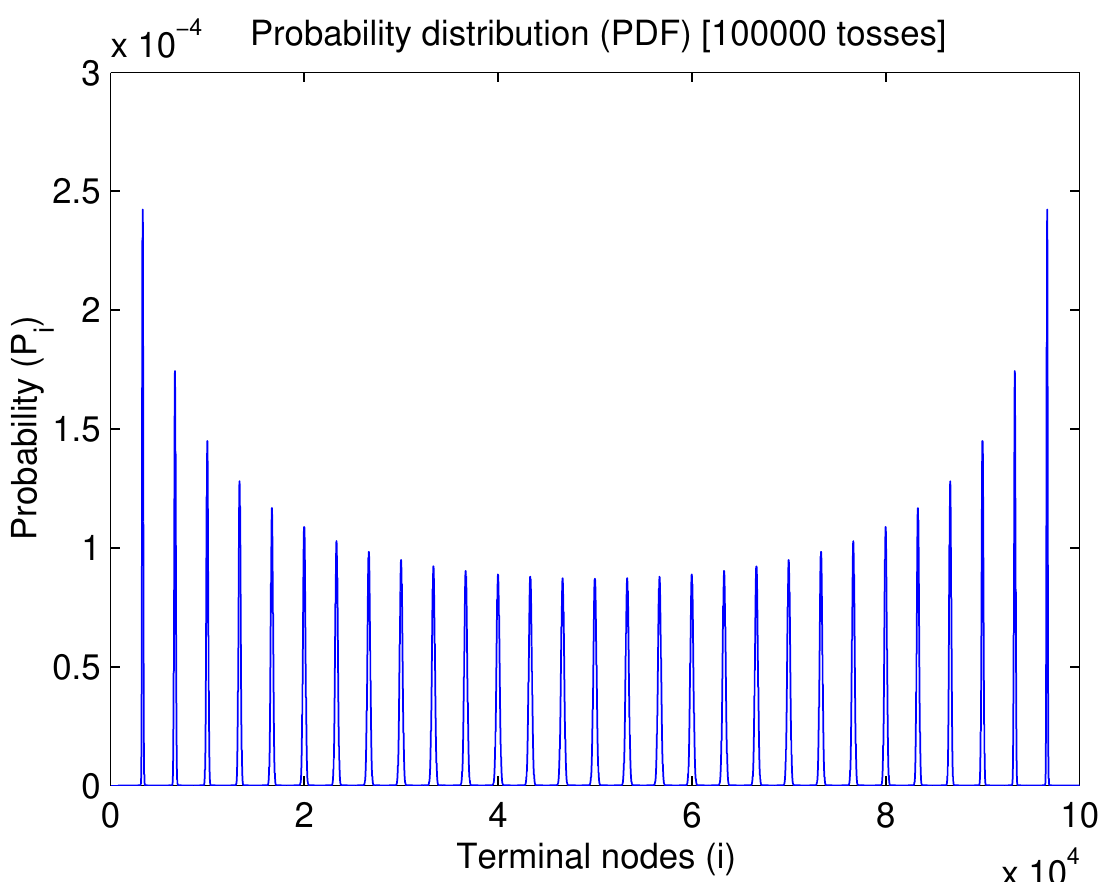}
\caption{The probability that heads occurs $m$ times in $t$ rounds
of the game with $q(p) = p$, assuming uniform initial wealth, for
(a) $t = 1000$ and (b) $t = 10^5$.  In (b) note that, although some
peaks appear higher than others, the total weight of each peak is
the same.} \label{binomialWealth}
\end{center}
\end{figure}

It is possible to compute the distribution of wealth after $t$ steps
in closed form for the purely subjective case, $q(p) = p$.  The
probability that heads occurs $m$ times in $t$ steps is a sum of
binomial distributions, weighted by the initial wealths $w_i^{(0)}$
of the players,
\[
P_m^{(t)} = \sum_{j=1}^N
\w{j}{0}\left[{\binom{t}{m}}s_j^m(1-s_j)^{t-m}\right] \,,
\]
and the corresponding wealth of player $i$ is
\[
\w{i}{t}= \D \frac{s_i^m(1-s_i)^{t-m}\w{i}{0}}{\sum_j
s_j^m(1-s_j)^{t-m}\w{j}{0}}\,.
\]
When the initial wealths are evenly distributed among the players,
no player has an advantage over any other.  However, as soon as the
first coin toss happens, the distribution of wealth becomes uneven.
Wealthier players have an advantage because they have a bigger
influence on the odds, so the coin tends to acquire a bias
corresponding to the strategies of the dominant (i.e.\ initially
lucky) players.  Figure~\ref{binomialWealth} shows the probability
$P_m^{(t)}$ for $t = 1000$ and $t = 10^5$. After $1000$ steps the
binomial distributions are still strongly overlapping, and there is
still a reasonable chance to overtake the winning strategy.  After
$10^5$ steps, however, the bias of the coin has locked onto an
existing strategy $s_i$, due to the fact that this strategy has
almost all the wealth.  Once this happens, the probability that this
will ever change is extremely low.

We now explain the peculiar bias dynamics observed for
the multi-modal map $q(p) = 3p \mod 1$ in
Figure~\ref{biasDynamics}(e), in which the bias of the coin
oscillates wildly between $0$ and $1$.  Through the
passage of time the wealth becomes concentrated on strategies near
either $s = 1/3$ or $s = 2/3$, corresponding to the
discontinuities of $q(p)$.  Suppose, for example, that $p = \sum_i
w_i s_i$ is slightly greater than $1/3$, where the $q(p)$ map is
close to zero.  This causes a transfer of wealth toward strategies
with smaller values of $s$ until $p = \sum_i w_i s_i < 1/3$.  At
this point the bias of the coin flips because $q(p)$ is now close to
one and the transfer of wealth reverses to favor strategies with
higher values of $s$.  Due to fluctuations in the outcomes of the
coin toss this oscillation is not completely regular.   It continues
indefinitely, even though with the passage of time wealth becomes
concentrated more and more tightly around $s = 1/3$.  A
similar process occurs if the first coin tosses cause convergence around $s = 2/3$.  We discuss the initial convergence around $s =1/3$ or $s = 2/3$ in more detail in the next section.

\section{Rational players and Nash equilibria}

So far we have studied the evolution with a fixed set of strategies.
What happens when we instead consider rational players?  What are
 the Nash equilibria?  We will see that while the general
situation here is quite complicated, we can nonetheless get some
 analytic insight into the attractors of the game with fixed strategies.
Furthermore, we will argue that if one is interested in surviving to
dominate the game, the choice of utility function is not arbitrary.

The first question that must be addressed is, ``What would rational
players reasonably optimize?". The answer is not obvious.  For
example, suppose a player maximizes the expected payoff on the next
 step. Assuming there is higher expectation for the coin to yield heads, the
strategy that maximizes the expected payoff is $s_i = 1$, i.e.\
betting all the player's wealth on heads.   However, this gives a
 non-zero probability of bankruptcy, and in the long run guarantees that
this player's wealth will asymptotically be zero. This illustrates
the need for risk aversion, and the need to look ahead more than one
move.  In general looking ahead is very complicated, due to the fact
that each player's move affects not only the player's expected
wealth on the next step, but also the future bias of the coin, and
hence the future wealths of all other players.

For the case $q = \text{constant}$ it is possible to show that the
strategy that asymptotically accumulates all the wealth maximizes
the expected log-return \cite{CoverThomas:book}
\begin{eqnarray*}
\E\r{i}{t+1} & = & \E \log \frac{\w{i}{t+1}}{\w{i}{t}} \\
& = & \E\log\frac{\pi_i}{\w{i}{t}} = \sum_l q_l \log \frac{s_{il}}{p_l}.
\end{eqnarray*}
For $q = \text{constant}$ and fixed strategies, maximizing
the log-return repeatedly for one step is equivalent to maximizing it over a long
time horizon.  For a general reality map $q(p)$, however, this is no
longer sufficient; it is easy to produce examples for which two step
optimization produces different results than single step
optimization, due to the effect of changes in the strategy's wealth
on the future bias of the coin.  For the case where the strategy's
wealth is negligible, however, the situation is greatly simplified:
It is possible to show that maximizing the one step log-return at
each step is equivalent to maximizing the log-return over many steps
\cite{Cherkashin04}.  In a game with many players, each of whom has
small wealth, maximizing the log-return for the next step may be a
good approximation to the optimal strategy.  As we have already
seen, acquiring a lead in the early stages of the game gives an
advantage that tends to persist later on.

The expected log return for one step can be written
\[
r(s) = q\log\frac{s}{p} + (1-q)\log\frac{(1-s)}{(1-p)},
\]
where as before $q = q_1$, $s = s_1$, $w = w_1$, etc.  The first
derivative is
\begin{multline*}
\frac{dr}{ds} = q'w\bigg[\log s - \log p - \log(1-s) + \log(1-p)\bigg] \\
 + q\left(\frac{1}{s} - \frac{w}{p}\right) + (1-q)\left(-\frac{1}{1-s} + \frac{w}{1-p}\right),
\end{multline*}
where $q' = dq/dp$.  A sufficient condition for $dr/ds = 0$ is $s =
q = p$.  In this case the second derivative is
\[
\frac{d^2 r}{ds^2} = \frac{1-w}{s(1-s)}\bigg[2wq'-(1+w)\bigg].
\]
For the map $q_\alpha$ the strategy $s = 1/2$ is a local maximum
when $\alpha < (1 + w)/(2w)$.  Providing this condition holds, this
implies that this strategy is what we will call a \textit{myopic log
Nash equilibrium}, i.e., it is a strategy that, when played against
itself, gives the best possible expected log return for the next
round of the game.   The myopic log Nash equilibria depend on the
reality map $q(p)$, but in general they also depend on the wealth of
the players, so that they can be dynamic, shifting with each coin
toss.   For example, when $\alpha <  1$, $s = 1/2$ is always a
myopic log Nash equilibrium, but when $\alpha > 1$ the myopic log
Nash equilibria strongly depend on the wealth of the players.  This
makes long-range optimization difficult.

\begin{figure}[htbp]
\begin{center}
\includegraphics[width=0.35\textwidth]{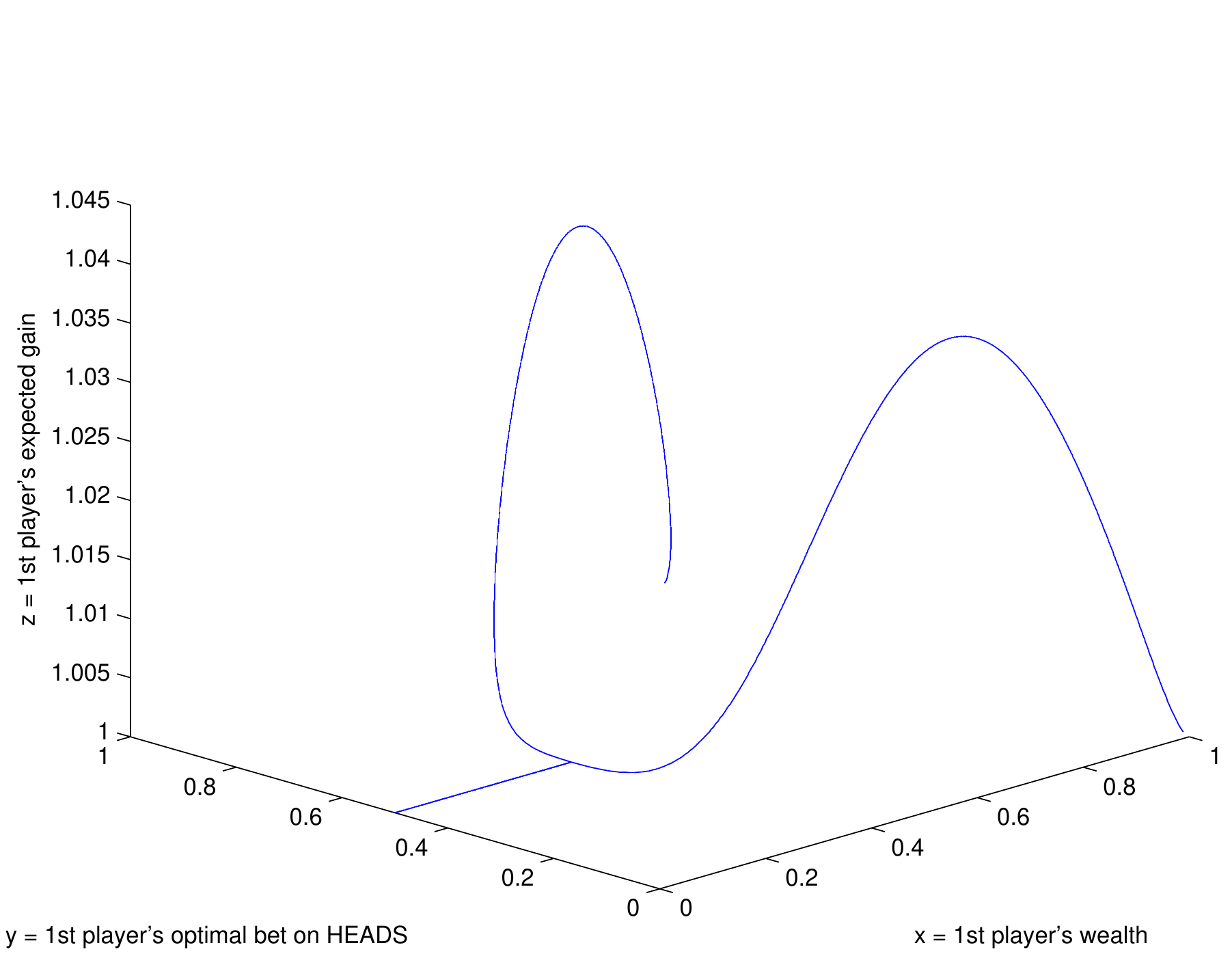}
\caption{The expected log-return $r(s)$ for a player (arbitrarily labeled ``first") using a
 myopic log Nash strategy against a player using a fixed strategy $s = 1/2$, where $q(p) =
q_\alpha$ with $\alpha = 2$.  This is plotted against the first
player's wealth on the $x$ axis and the first player's optimal bet
on heads on the $y$ axis.} \label{nashEq}
\end{center}
\end{figure}

 This is illustrated in
Figure~\ref{nashEq}, where we show the expected log return for a
player (arbitrarily labeled ``first") playing a myopic log Nash strategy against another (second) player using a fixed strategy with $s =1/2$.  When the first player's wealth is low, $1/2$ is the optimal
strategy, and it is a myopic Nash equilibrium. As the first player
gains in wealth two strategies on either side of $s = 1/2$ become
 superior; as wealth increases, these strategies become more and more separated from $s =
1/2$, and in the limit as $w \to 1$ the optimal strategy is either
$s = 0$ or $s =1$.

 In the limit as the first player's wealth gets large the possible myopic log Nash equilibria do a good job of predicting the attractors
of the objective dynamics shown in Figure~\ref{biasDynamics}.  For
 the self-defeating case, $q(p) = 1 - p$, there is a unique myopic log Nash
 equilibrium at $s = 1/2$, corresponding to the unique attractor for $q$.
For the self-reinforcing case with $\alpha < 1$ this is also true.
 For the identity map $q(p) = p$ the entire interval $0 \leq s \leq 1$ is a myopic log Nash
 equilibrium.  When $\alpha \gg 1$ either $s
= 0$ or $1$ can be myopic log Nash equilibria.  In each case these
correspond to the attractors of the wealth dynamics.

The multimodal map $q(p) = 3p \mod 1$ is interesting because all the
intersections with the identity are local minima for the expected
log return.  Instead, the system is attracted to the discontinuities
of the map at $s = 1/3$ and $s = 2/3$.  It is as if one can think of
the discontinuities of the map as being connected (with infinite
slope), creating intersections with the identity that yield local
maxima of the log return.  While we do not have a formal method of
proving this, we gave an intuitive explanation for how this
stability comes about in Section~\ref{wealthDynamics}.

\section{Efficiency}

As the game is played the reallocation of wealth causes the
population of players to become more efficient in the sense that
there are poorer profit opportunities available for optimal players.
This is analogous to financial markets, where wealth reallocation
due to profitable vs.\ unprofitable trading has long been
hypothesized to result in a market that is efficient in the sense
that all excess profit-making potential has been removed.  In
financial economics efficiency is taken as a postulate; there is no
theory that guarantees that markets out of equilibrium will always
be attracted to a perfectly efficient market equilibrium, and no way
to quantitatively measure the inefficiency of a market when it is
 out of equilibrium.  Our game
provides a simple setting to study the approach to efficiency in an
out-of-equilibrium context.

We can measure the inefficiency of our game based on the returns of
 what we will call a \textit{rational $\epsilon$ player}.  This player knows the
strategies of all other players, and pursues an optimal strategy
that maximizes her expected log returns.  This player has
infinitesimal wealth $\epsilon$, so that her actions have a
negligible effect on the outcome of the game.  In the purely
objective setting where $q = \text{constant}$ the approach to
efficiency is guaranteed by the fact that the wealth dynamics are
formally equivalent to Bayesian updating, implying all the wealth
converges on the correct hypothesis about the bias of the coin. For
more general settings this is no longer obvious, as there is no
longer such a thing as an objectively correct hypothesis.

\begin{figure}[htbp]
\begin{center}
\includegraphics[width=0.5\textwidth]{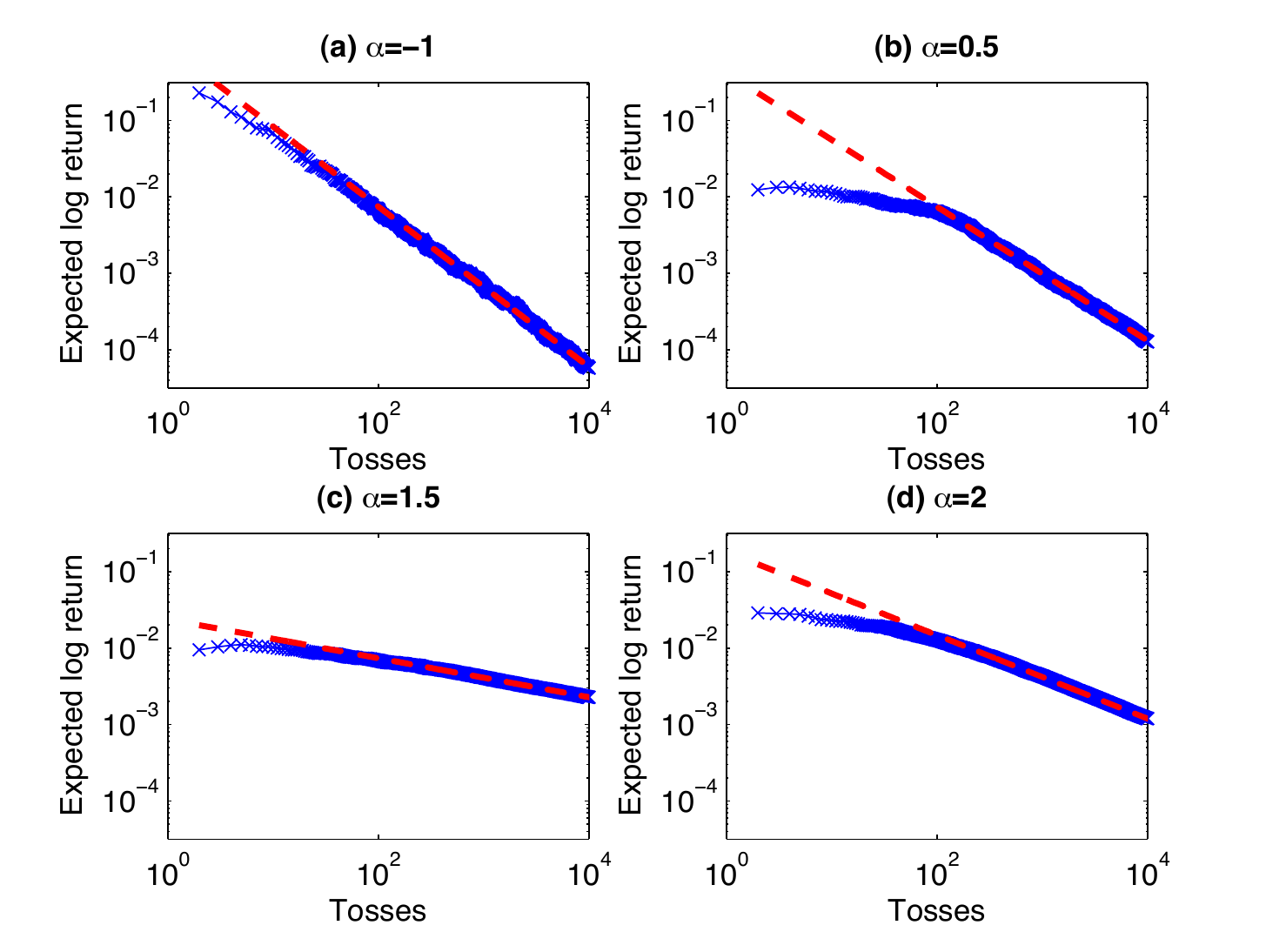}
\caption{The inefficiency of the game is measured by the log
returns of an optimal $\epsilon$ player, shown as a function of time using
$N = 3000$ players.  Plots are in double logarithmic scale.  (a)
$q(p) = 1-p$;  (b) $\alpha = 0.5$;  (c) $\alpha = 1.5$; (d) $\alpha
= 2.0$.} \label{efficiency}
\end{center}
\end{figure}

We have studied the approach to efficiency numerically for a variety
of different reality maps as shown in Figure~\ref{efficiency}.  To
damp out the effect of statistical fluctuations from run to run we
take an ensemble average by varying the random number seed.  For
$q(p) = p$ we find that the inefficiency is essentially zero at all
 times.  In every other case we find that the
efficiency is a decreasing function of time, asymptotically
converging as a power law $t^{-\gamma}$ with $0 \le \gamma \le 1$.
For the self defeating case $q(p) = 1 - p$ we observe $\gamma
\approx 1$; for other values of $\alpha$ we observe $\gamma < 1$.
For example, for $\alpha = 0.5$, $\gamma \approx 0.6$, for $\alpha =
1.5$, $\gamma \approx 0.25$, and for $\alpha = 2$, $\gamma \approx
0.5$.  For maps close to the purely subjective case, the
inefficiency is initially quite small but convergence is
correspondingly slow.   For example, compare \ref{efficiency}~(a)
and (b).  The self-defeating case is initially much more
inefficient than the mildly self-reinforcing case, but by $t =
10,000$ the situation is reversed.  The rate of convergence in
efficiency reflects the slow convergence in the bias of the coin to
its fixed point attractor.\footnote{We have performed simulations
with different numbers of agents and find that domain of validity of
the power law scaling is truncated for small $N$ (e.g. for $N
\approx 30$ it extends only to roughly $t \approx 1000$). The length
of validity of the power law scaling in time increases with $N$.
Thus there is a finite size effect, indicating that the power law
scaling is exactly valid only in the limit as $N \to \infty$ and $t
\to \infty$.}

\section{Increasing returns}

When the reality map for a game of chance is purely objective a
player can't manipulate the outcome unless she cheats.  In contrast,
for more general reality maps, under some circumstances the player
can use the subjective dependence of $q(p)$ to manipulate the
objective odds to her advantage.  This manifests itself as
 increasing returns to scale\,---\,as the player acquires more wealth,
the ability to manipulate the odds increases, and the expected
return also increases.

\begin{figure}[htbp]
\begin{center}
\includegraphics[width=0.2\textwidth]{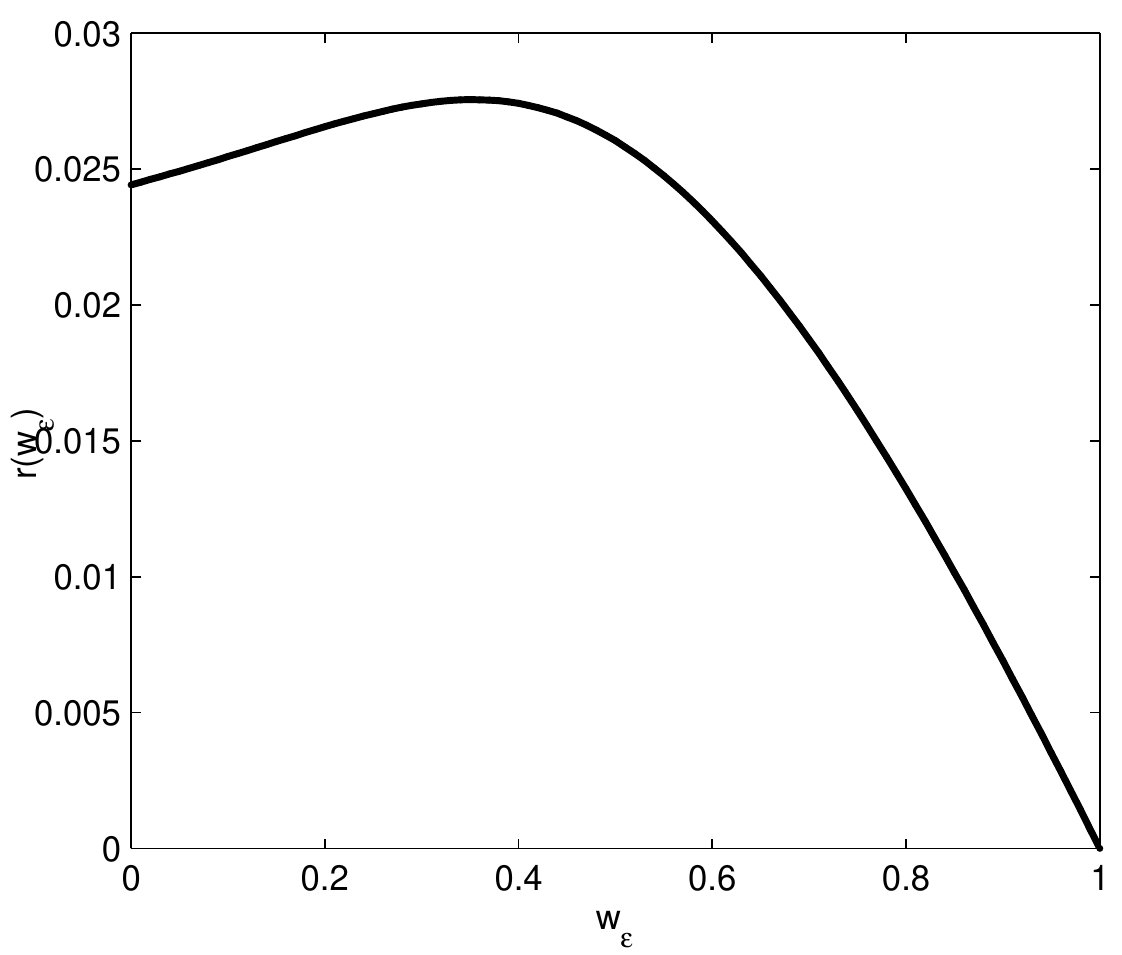}
\includegraphics[width=0.2\textwidth]{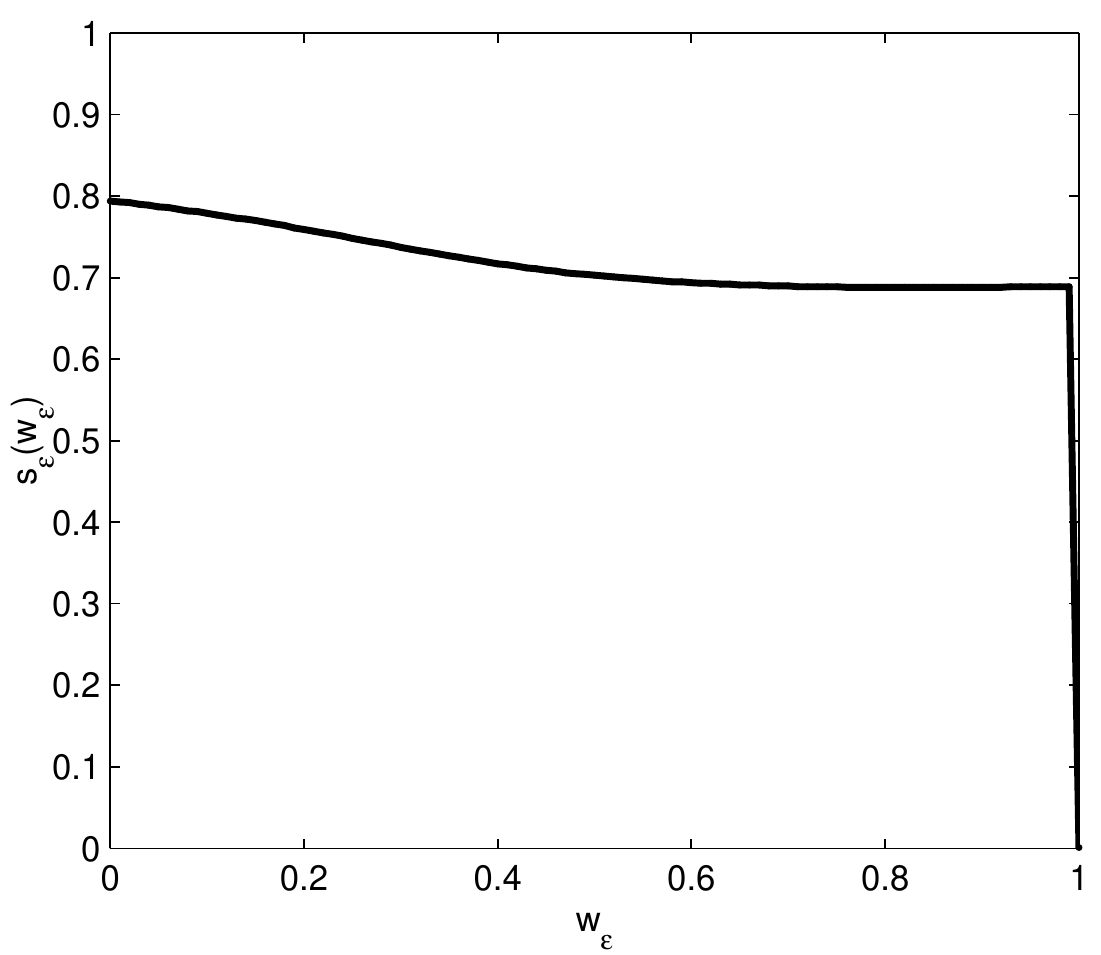}
\includegraphics[width=0.2\textwidth]{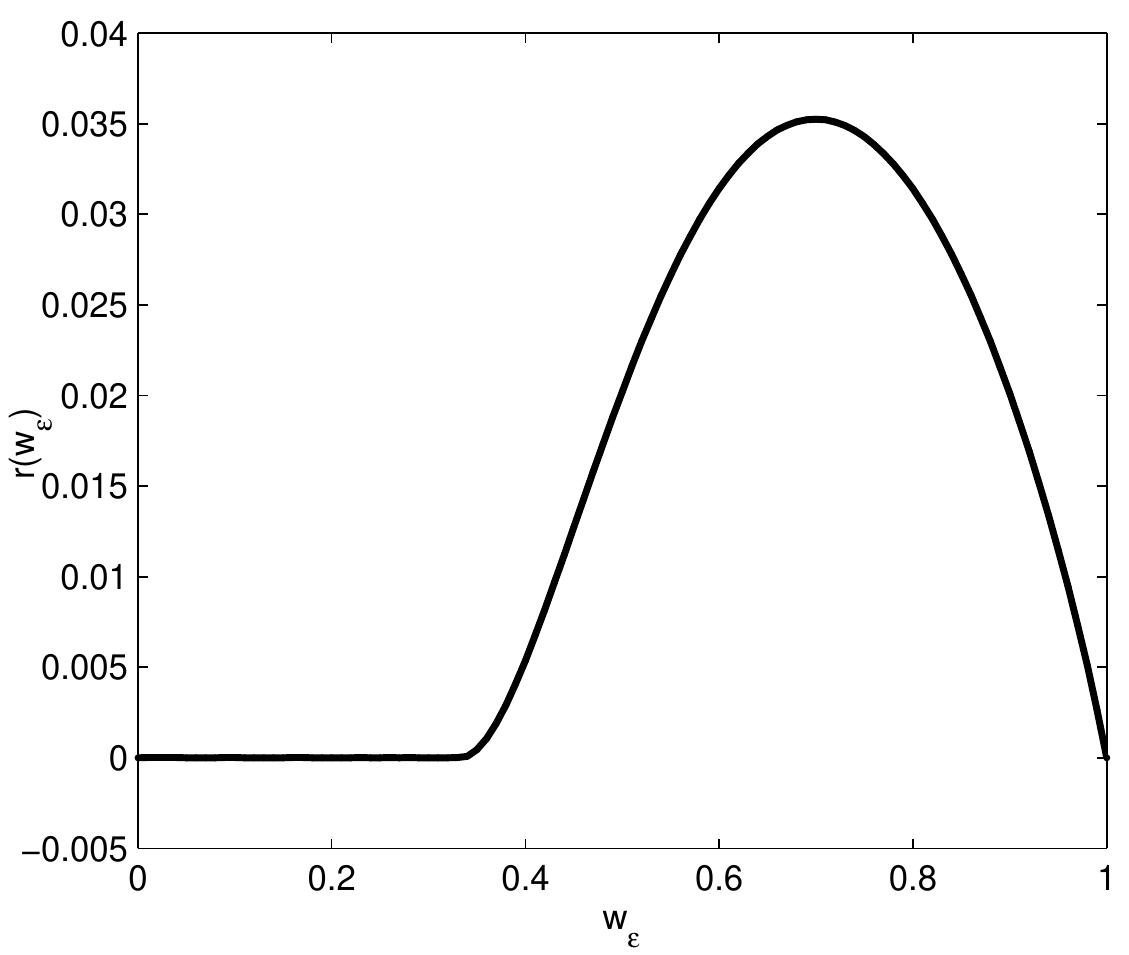}
\includegraphics[width=0.2\textwidth]{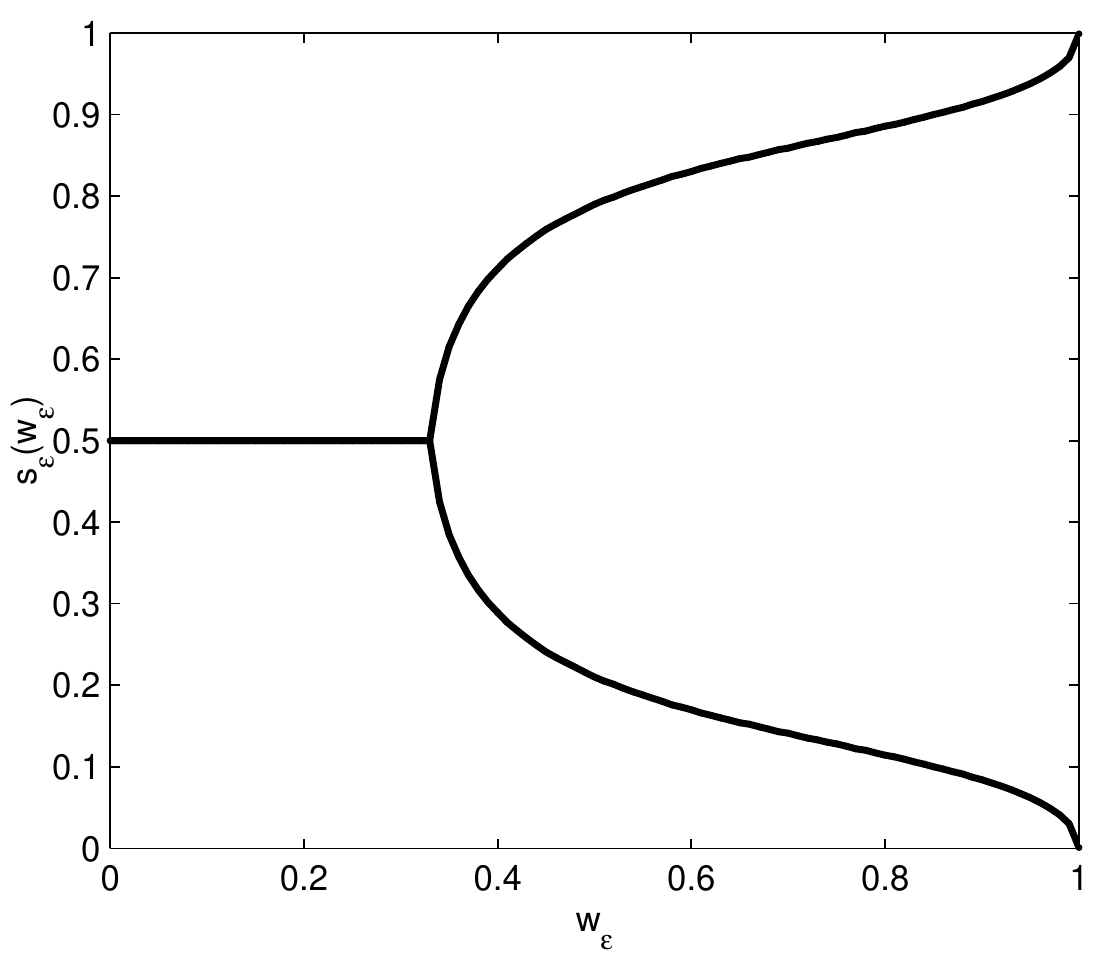}
 \caption{The left column shows the expected log returns $r(w)$ for an optimal player and the right column shows the corresponding optimal strategies $s(w)$, both as a function of her wealth $w$.
In the top row $\alpha = 1/2$ and $t = 7$, where $w_i^{(7)}$ is produced by starting at $\w{i}{0}=1/N$ and generating 7 heads, where $N=29$.  In the bottom row $\alpha=2$, $t = 0$, and $w_i^{(0)} =1/N$.}
\label{increasingReturns}
\end{center}
\end{figure}

To measure this we study a rational strategy that exploits its
complete knowledge of the strategies of the other players to
maximize the expected log-return for the next step.  We start with
given wealth assignments for the players and vary the fraction of
 the wealth for the rational player vs.\ the other players. Under some
circumstances, which depend both on $q(p)$ and the distribution of
wealth, for self-reinforcing reality maps we find that the returns
are an increasing function of the wealth of the rational player.
Two examples are given in Figure~\ref{increasingReturns}.

\section{Summary}

We have introduced a very simple evolutionary game of chance with the nice property that one can explicitly study the influence of
the player's actions on the outcome of the game.  By altering the
reality map $q(p)$ it is possible to continuously vary the setting
from completely objective, i.e.\ the odds are independent of the
players' actions, to completely subjective, i.e.\ the odds are
completely determined by the players' actions.

This is an evolutionary game in the strong sense:  Only one player
survives to have non-negligible wealth.  Our results suggest that
the myopic maximization of expected log returns is a fairly good
survival strategy, certainly much better than simply maximizing
returns.  However, in contrast to the purely objective case, it is
provably not an optimal strategy.  This is due to the complications induced
by the feedback between the success of the players and the objective
reality as reflected in the bias of the coin.


It has long been known that subjective effects can play an important
role in games, causing problems such as increasing returns and
lock-in to a particular outcome due to chance events.  This model
shows that the existence of subjective effects alone are not enough.
Instead, for most of the properties we study here, such as selecting
between alternative equilibria or increasing returns, we need the
self-reinforcing effects to be sufficiently strong to destabilize
the objective dynamics.  There is a competition between the
stabilizing force of wealth concentration and the destabilizing
properties of $q(p)$ when $q' > 0$.  This game shows how these
effects become steadily stronger as the self-reinforcing nature of
the reality map increases.  It also shows that these effects are
generally complicated and wealth dependent, even in this simple
 situation.  The myopic log Nash equilibria are strongly wealth dependent.  Since
 wealth evolves through time, the myopic log Nash equilibria also evolve in a
time dependent manner.

This game provides a setting in which to study the progression
toward efficiency in an out-of-equilibrium context.  We have
introduced a notion of efficiency that closely resembles arbitrage
efficiency in financial markets.  We always observe a progression
toward efficiency, except for the purely subjective case, in which
 it appears that any configuration of player strategies automatically produces an
efficient market.  This isn't surprising, since in the purely
 subjective case there is no preferred strategy.  In every other case, as wealth is reallocated,
the game becomes more efficient, in the sense that there are fewer
profit-making opportunities for skillful players.  For the examples
we observe that the inefficiency as a function of time
asymptotically decreases as a power law.

One might consider several extensions of the problem studied here.
 For example, one could study learning (see e.g.\ \cite{sato-akiyama-farmer:PNAS}).
Another interesting possibility is to allow more general reality
maps, in which $q$ is a multidimensional function with a
multidimensional argument that may depend on the bets of individual
players.  For example, an interesting case is to allow some players,
who might be called pundits, to have more influence on the outcome
than others.  It would also be very interesting to modify the game
so that it is an open system, e.g.\ relaxing the wealth conservation
condition and allowing external inputs.  This may prevent the
asymptotic convergence of all the wealth to a single player,
creating more interesting long-term dynamics.

\section{Acknowledgements}

We would like to thank Jonathan Goler for his help with simulations
in the early days of this project, and to Michael Miller for
reproducing the results and clarifying the scaling in
Figure~\ref{efficiency}.  JDF would like to thank Barclays Bank and National Science Foundation grant 0624351 for support.  Any opinions, findings, and conclusions or recommendations expressed in this material are those of the authors and do not necessarily reflect the views of the National Science Foundation.

\bibliography{pnas}
\bibliographystyle{plain}

\end{document}